\documentclass[a4paper, 10pt, conference]{ieeeconf}      

\IEEEoverridecommandlockouts                              
\usepackage{graphics} 
\usepackage{epsfig} 
\usepackage{mathptmx} 
\usepackage{times} 
\usepackage{amsmath} 
\usepackage{amssymb}  
\usepackage{subfigure}

\newtheorem{defn}{Definition}
\newtheorem{thm}{Theorem}

\newtheorem{lem}{Lemma}

\newcommand{\bmf}[1]{\mbox{\boldmath$#1$}}

\newcommand{\robsin}{\textsc{Robsin}}

\title{\LARGE {\bf Methods for robust PID control}\\
\vspace{0.2cm}
}

\author{Naim~Bajcinca
\thanks{N. Bajcinca is with the Max-Planck Institute for Dynamics of Complex Technical Systems in
Magdeburg, Germany. The work presented here was done while working with the group of Prof.~J.~Ackermann at German Aerospace Research Center (DLR) in Oberpfaffenhofen, Germany.}}

\begin{document}

\maketitle
\thispagestyle{empty}
\pagestyle{empty}

\begin{abstract}
A comprehensive theory for robust PID control in continuous-time and discrete-time domain is reviewed in this paper. For a given finite set of linear time invariant plants, algorithms for fast computation of robustly stabilizing regions in the ($k_P,~k_I,~k_D$)-parameter space are introduced. The main impetus is given by the fact that non-convex stable regions in the PID parameter space can be built up by convex polygonal slices. A simple and an elegant theory evolved in the last few years up to a quite mature level. \end{abstract}

\section{INTRODUCTION}
\label{sec:Introduction} It is a well-known fact that by far the most applied control law for SISO systems
in nearly all industries (process control, motion control, aerospace etc) is the PID control. No other controller can compete to PID when it comes to performance per simplicity of the control structure, this being the reason for its absolute dominance in the practice. Traditionally, tuning of a PID controller has been the overwhelming usability approach in research and applications. The design technique presented here is in some sense an opposing one. We want to compute the total set of PID-stabilizers. While it turns out to be interesting in theoretical terms, its impact on practical applications is difficult to predict. Advanced software tools based on this technology (e.g. \robsin, \cite{bajc:cca2004}) suggest to the user a 3D-region in ($k_P,~k_I,~k_D$)-parameter space, where he can select a controller from. By doing so, he would additionally have an idea how robust (i.e. how far from instability) his design is. A further good news is that the same technique applies for time-delay systems and in discrete-time domain.

This paper covers theoretical design issues only. It has been shown that the stabilizing region for continuous-time PID ($=k_P+k_I/s+k_D s$) controllers is defined by a set of convex polygonal slices normal to $k_P$ axis in the $(k_P,~k_I,~k_D)$ parameter space, see \cite{bhatta:book2000}, \cite{kaes:ecc2001}, \cite{bajc:auto:2006}. The method followed by the author uses the decoupling effect of PID parameter space at singular frequencies, \cite{bajc:auto:2006}. Thereby the characteristic polynomial decouples into two frequency parameterized equations, one involving $k_I$ and $k_D$, and another one with $k_P$ only. As a consequence, the computation of all stable PID controllers may be divided into two subproblems: assertion of stable intervals of parameter $k_P$ (so-called $k_P$-problem), and detection of stable polygons on the plane ($k_I,~k_D$) for a given $k_P$. In \cite{bajc:ifac2002} and \cite{bajc:ecc07} the design approach was transferred to discrete-time systems, and in \cite{bajc:tds2004} to time-delay systems.

A general algorithm for automatic detection of stable polygons is based on the analysis of the motion of eigenvalues in the vicinity of the singular frequencies. This algorithm was first presented in \cite{bajc:med2001}. It detects the inner polygons and selects the one with maximal stable eigenvalues, which is finally checked for stability. It is important to emphasize, that this algorithm can be applied equally well for PID quasi-polynomials describing the PID control loops for time-delay systems.

Preserving simplicity has been a basic motivation in searching for a solution to the $k_P$-problem. A simple necessary condition was firstly understood in \cite{bajc:acc2005}. It turns out that for a given plant a sufficient number of singular frequencies must be available for its stabilization. Since $k_P$ uniquely determines the number of singular frequencies, one can directly read $k_P$-intervals from the $k_P$-plot, which may host stable polygons. This simple criterion can be extended to discrete-time domain and time-delay quasi-polynomials, for instance, see \cite{bajc:ecc07} and \cite{bajc:tds2004}.

The presented methods may be directly applied for the computation of the total robust set of PID controllers which stabilize a finite set of plants (multi-model). This is exactly what is meant by robust design of PID controllers.

In this paper methods for PID control in continuous- and discrete-time domain, as well as for time-delay systems, are reviewed. Therefore, we will have to switch between the $s$ domain (continuous-time) and $z$-domain (discrete-time) while presenting results. The paper is organized as follows. In Section~\ref{sec:problemDefinition} we postulate the stabilizability problem. Section~\ref{sec:veryBasicIdea} presents all steps of the design process, and simultaneously introduces the formulates technical problems, in a simple-case study. Theoretical fundamentals of the methods are introduced in Section~\ref{sec:basicDefinitions}. Automatic detection of convex stable polygons and $k-P$-problems are discussed in Section~\ref{sec:stablePolygons} and \ref{sec:kp-Porblem}, respectively. Finally, in Section~\ref{sec:TimeDelaySystems} the ideas are extended for time-delay systems. For illustration purposes, throughout the paper examples are provided. The reader may follow them by using the toolbox \robsin, which can be downloaded from \emph{http://www.robotic.dlr.de/robsin}.

This article primarily reviews the original work of the author. It has not been an intention to refer to the all work in the area. Still, the key contributions (and contributors!) to the theory are referred to and they read as follows. The role of convex polygons was firstly understood by Battacharyya and his co-workers, see the monograph \cite{bhatta:book2000}. Their derivation bases on Hermite-Biehler theorem. For computation of stable polygons they proposed linear programming techniques, but they did not really address the $k_P$-problem. Munro's computation is based on the real-axis intersections of the Nyquist plot, see, \cite{munro:2000}, \cite{munro:2003}. The relationship to singular frequencies was noticed firstly by Ackermann and Kaesbauer, see \cite{kaes:ecc2001}. Soylemez proposed a solution to the $k_P$-problem, however in author's opinion his criterion is not as simple and usable as the one presented in this article. The remainder theory reviewed here is fairly developed by the author, including the algorithms for automatic detection of stable polygons and solutions to the $k_P$-problem, as well as generalizations to discrete-time domain and time-delay systems.

\section{Problem definition}
\label{sec:problemDefinition}
\subsection{Discrete-time domain}~Consider a simple closed curve $\partial\Gamma~=\{z\mid
z=\tau(\alpha)+j~\eta(\alpha),~\alpha~\textrm{is a parameter}\}$, in $z-$plane,
which is symmetric to the real axis $\tau$ and can be expressed in
the form
\begin{equation}
    \label{eq:gammaeq}
    \partial\Gamma:~~F(\tau, \eta)=0.
\end{equation}
Let
\begin{equation}
\label{eq:basicEquation} p=A(z)Q(z,r_1,r_2,r_3)+B(z)=0
\end{equation}
be a three-term  algebraic equation in ($r_1,~r_2,~r_3$), with given polynomials $A$ and $B$
\begin{eqnarray}
    \label{eq:polynA}
    A(z)&=&a_0+a_1 z+\cdots+a_m z^m\\
    \label{eq:polynB}
    B(z)&=&b_0+b_1 z+\cdots+b_{n} z^{n},
\end{eqnarray}
and the second-order polynomial $Q$ in the form
\begin{equation}
\label{eq:QEquation}
Q=\delta_1(z)r_1+\delta_2(z)r_2+\delta_3(z)r_3.
\end{equation}

In this article we want to compute the set of all parameters $r_1,~r_2,~r_3$, s.t. the polynomial
(\ref{eq:basicEquation}) is $\Gamma-$stable, that is, all its eigenvalues
must lie within the $\Gamma$-region (enclosed by $\partial\Gamma$). Of main interest
are circles with center on the real axis $\tau$ and an arbitrary radius, which will be
referred to as $\Gamma-$circles. For discrete-time systems particularly important is the unity Schur-circle.

It may be easily shown that (\ref{eq:basicEquation}) describes the
characteristic equation of a feedback loop with a PID or a
three-term controller. Indeed, a discrete-time equivalent of the
PID controller has the transfer function
\begin{equation}\label{eq:PID}
K(z)=\frac{c_1 + c_2z + c_3z^2}{(z + z_1)(z - 1)}.
\end{equation}
Its structure follows in the quasi-continuous
consideration by applying the rectangular integration rule $(s
\rightarrow (z -1)/Tz)$ to the ideal PID controller $k_I/s + k_P +
k_Ds$, resulting in $z_1 = 0$, or by the trapezoidal integration
rule $(s \rightarrow 2(z - 1)/T(z + 1))$, resulting in $z_1 = 1$.
Also the realizable PID controller $k_I/s + k_P + k_Ds/(1 + T_1s)$
converts by the trapezoidal integration rule to the controller
structure (\ref{eq:PID}) with a pole at $z_1 = -(2T_1 - T)/(2T_1 +
T)$. Equation (\ref{eq:basicEquation}) covers also a three-term
controller with an arbitrary second order denominator polynomial
\begin{equation}\label{eq:threeTerm1}
 K(z)=\frac{n(z)(c_1 + c_2z + c_3z^2)}{d(z)}.
\end{equation}
For both control structures (\ref{eq:PID}) and
(\ref{eq:threeTerm1}), the polynomial $Q$ is of the form
\begin{equation}
\label{eq:SchurQ'Equationc1c2c3} Q=c_1+c_2 z +c_3 z^2.
\end{equation}
It is clear that (\ref{eq:QEquation}) and
(\ref{eq:SchurQ'Equationc1c2c3}) are connected by a linear transformation
\begin{equation}\label{eq:Tmatrix}
c=T~r,~~\textrm{det~}T~\neq 0,
\end{equation}
with $c=[c_1,c_2,c_3]^T,~~r=[r_1,r_2,r_3]^T$. The matrix $T$ is determined by the polynomials
$\delta_1(z),~\delta_2(z),~\delta_3(z)$.

\subsection{Continuous-time domain}~The pendant equation (\ref{eq:basicEquation}) for
a feedback-loop with a PID controller in continuous-time domain is
\begin{equation}\label{eq:basicCharacteristicPolynomial}
    p = {A}(s)( k_I+ k_P s+  k_D s^2)+ {B}(s)
\end{equation}
where polynomials $A(s)$ and $B(s)$ are as in (\ref{eq:polynA}) and (\ref{eq:polynB}). As in the discrete-time case, we want to compute the set of all parameters $k_P$, $k_I$ and $k_D$ for which the polynomial (\ref{eq:basicCharacteristicPolynomial}) is Hurwitz-stable, i.e. all of its eigenvalues lie on the left-hand side of the imaginary axis $s=j\omega$. In other words, here $\partial\Gamma=\{j\omega:~\forall \omega\in \mathbb{R}\}$. In view of the definition (\ref{eq:basicEquation}), the polynomial $Q$ is also of the form (\ref{eq:SchurQ'Equationc1c2c3})
\begin{equation}
\label{eq:HurwitzQ} Q=k_I+k_P s +k_D s^2.
\end{equation}
Obviously discrete-time consideration is more general, with $z,\alpha,r_1,r_2,r_3$ corresponding to $s,\omega,k_I,k_D,~k_P$, respectively.

\section{The very basic idea}
\label{sec:veryBasicIdea}
\begin{figure}[h]
\centering
\subfigure[The $k_P$-plot]{
\label{fig:kpplot}
\includegraphics[width=.4\textwidth]{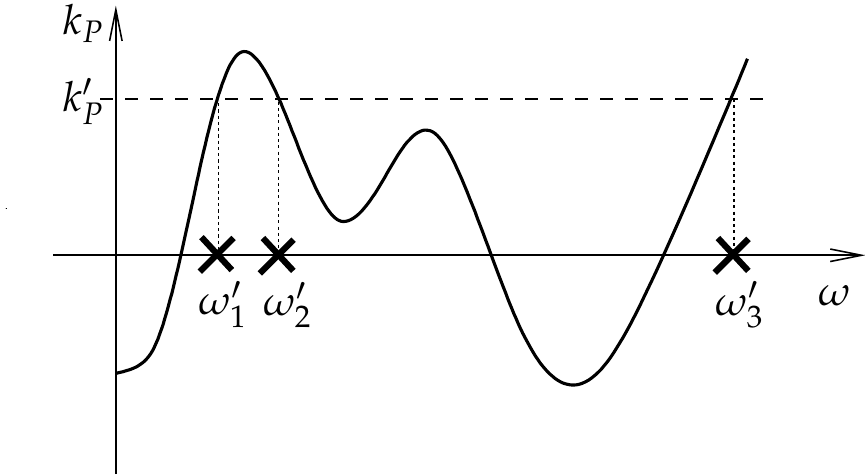}}\\
\subfigure[Mapping of singular frequencies into the ($k_I,k_D$)-plane]{
\label{fig:kikdplane}
\includegraphics[width=.4\textwidth]{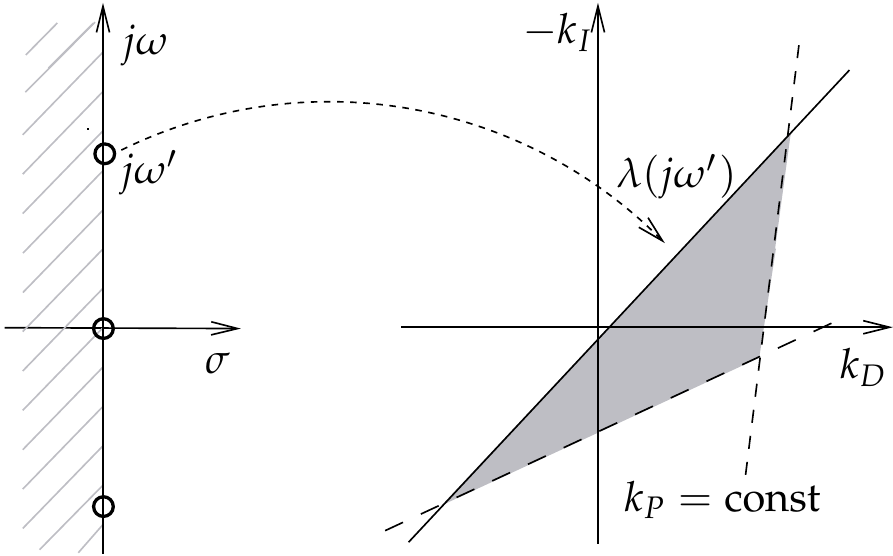}}\\
\vspace{.1in}
\caption{Illustrating the basic algorithm for computation of stable PID controllers in continuous-time domain}
\label{fig:basicIdea}
\end{figure}

Consider the special case with $A(s)=1$ in (\ref{eq:basicCharacteristicPolynomial}). Substitution $s=j\omega$, decouples the latter into two equations
\begin{equation}\label{eq:basicIdea}
k_I -\omega^2 k_D=-R_B, ~~k_P=-I_B/\omega
\end{equation}
where $R$ and $I$ stand for the real and imaginary part of polynomial
$B(s)$ at $s=j\omega$. Notice that PID parameters appear decoupled in tow equations. Computation of the stable PID region should be quite obvious for this case study. (1): First, for a fixed parameter $k_P=k'_P$ solve for the frequencies $\omega'$ from the second equation in (\ref{eq:basicIdea}), representing the {$k_P$-plot} and shown in Fig.~\ref{fig:kpplot}. Such frequencies are known as \emph{singular frequencies}. (2): Map all singular frequencies using the first equation in (\ref{eq:basicIdea}) into the $(k_I,k_D)-$plane as shown in Fig.~\ref{fig:kikdplane}, and compute the stable polygons (gray area in Fig.~\ref{fig:basicIdea}). Hereby, each pair of singular frequencies $\pm j\omega'$ maps to a straight boundary line. Note that, as parameters $k_I$ and $k_D$ are varied, with $k_P$ kept fixed, the eigenvalues of (\ref{eq:basicEquation}) can cross over imaginary axis $j\omega$ at singular frequencies only. This procedure is repeated for other $k_P$ parameters, which yield other stable polygons. Thus a tomographic 3-D picture results, as that in Fig.~\ref{fig:3DsingLinesExample1}.

In the sequel, we want to generalize the above algorithm and provide solutions to the problems: [P1] For what values of the parameter $k_P$ stable polygons should be searched for - indeed one would like to exclude \emph{a priori} $k_P$-intervals with no stable polygons. This problem is referred to as the  \emph{$k_P$-problem}. [P2] For a given set of boundary lines, how to automate the computation of the stable polygons.

\section{Basic definitions and theorems}
\label{sec:basicDefinitions}
Let $H$ and $G$ represent the real and imaginary part of the
characteristic  polynomial $p(z,r_1,r_2,r_3)$ in
(\ref{eq:basicEquation}) on $\partial\Gamma$.

{\it
\begin{defn}\label{def:SingulaeresGammaGebiet} $\Gamma$ is
said to be singular with respect to parameters $r_1$ and $r_2$ in
(\ref{eq:basicEquation}) if
\begin{equation}\label{eq:RangebdingungFuerSingulaereGammaGebiete}
   \textrm{Rank}\frac{\partial{}(H,G)}{\partial{}(r_1,r_2)}=1\qquad \textrm{for
  all } z\in{}\partial\Gamma{}.
 \end{equation}
\end{defn}
} The latter equation is referred to as the \emph{rank-condition}. From now on, we only consider $\Gamma$s which satisfy (\ref{eq:RangebdingungFuerSingulaereGammaGebiete}). A zero of the polynomial (\ref{eq:basicEquation}) that
fulfills (\ref{eq:RangebdingungFuerSingulaereGammaGebiete}) is referred to as
\emph{singular (or critical) frequency}.

{\it
\begin{defn}\label{def:decouplingFunction}
A function $E_{\Gamma}(z)$ defined as
\begin{equation}\label{eq:decouplingFunction}
Q(z,r_1,r_2,r_3)=E_{\Gamma}(z)~q(z,r_1,r_2,r_3)
\end{equation}
with
\begin{equation}\label{eq:Iq1}
I_q=r_3 g_3(\alpha)+g_0(\alpha),~~~\alpha\in [a,~b]
\end{equation}
where $I_q$ stands for the imaginary part of $q$, will be referred to as decoupling function of $Q$ on
$\partial\Gamma$.
\end{defn}}
In other words, by introducing $E_{\Gamma}$, a function $q$ is extracted from $Q$, whose imaginary part depends on one parameter $r_3$ only. It can be easily checked that if $\Gamma$ satisfies (\ref{eq:RangebdingungFuerSingulaereGammaGebiete}), then
\begin{equation}\label{eq:derr1r2}
\frac{\partial I_q}{\partial r_1}=0 \Leftrightarrow \frac{\partial
I_q}{\partial r_2}=0,~~~\forall z\in \partial\Gamma.
\end{equation}
Furthermore, two trivial decoupling functions on $\partial\Gamma$  of $Q$ in (\ref{eq:QEquation}) are
\begin{equation}\label{eq:decouplingFuntion}
E_\Gamma(z) = \delta_1(z),~~~E_\Gamma(z) = \delta_2(z).
\end{equation}
Using these two facts, the next statement follows directly.
{\it
\begin{thm}\label{thm:decoupling}Consider the function
\begin{equation}\label{eq:Fz}
F(z):=\frac{p(z)}{A(z)E_\Gamma(z)}.
\end{equation}
The equation $F(z)=0$ for $z\in\partial\Gamma$ decouples the parameters
$r_1, r_2$ and  $r_3$ into two equations
\begin{eqnarray}
\label{eq:singFreqsLines}
r_1 h_1(\alpha)+r_2 h_2(\alpha)+h_0(\alpha)&=&0,\\
\label{eq:singFreqs}r_3 g_3(\alpha)+g_0(\alpha)&=&0.
\end{eqnarray}
\end{thm}}
Note that the latter equations reveal decoupling of the parameter space ($r_1,r_2,r_3$). They represent generalizations of the simple equations (\ref{eq:basicIdea}).

Finally, let $\{p_{\nu}\}$ represent a finite set of polynomials
of the form (\ref{eq:basicEquation}), i.e.
\begin{eqnarray}
    \label{eq:basicEquationFamily}
p_{\nu}=A_{\nu}(z)Q(z,r_1,r_2,r_3)+B_{\nu}(z).
\end{eqnarray}
For instance, this may be a multi-model of a continuum of plants or Kharitonov polynomials of an interval uncertainty. It can be simply proven that the rank-condition (\ref{eq:RangebdingungFuerSingulaereGammaGebiete}) does not depend on the polynomials $A_{\nu}(z)$ and $B_{\nu}(z)$. Hence, a singular $\Gamma$ is completely determined by the polynomial $Q$ in (\ref{eq:QEquation}). The polynomials $A_{\nu}(z)$ and $B_{\nu}(z)$ rather determine the singular frequencies on $\partial\Gamma$.

\subsection{Hurwitz-stability}
Consider the Hurwitz-region $\partial\Gamma~=\{j\omega:~\omega\in \mathbb{R}\}$. Then condition (\ref{eq:RangebdingungFuerSingulaereGammaGebiete})
\begin{equation}\label{eq:rankConditionHurwitz}
\textrm{Rank}\frac{\partial{}(H,G)}{\partial{}(k_I, k_D)}=1
\end{equation}
is satisfied everywhere on the imaginary axis, since
\begin{eqnarray}\label{eq:H}
H&=&R_A k_I-\omega^2 R_A k_D-\omega I_A k_P +R_B\\
\label{eq:G}
G&=&I_A k_I-\omega^2 I_A k_D +\omega R_A k_P +I_B.
\end{eqnarray}
Referring to equations (\ref{eq:decouplingFuntion}) and (\ref{eq:HurwitzQ}), a simple choice for the decoupling function is $E_\Gamma(z) = 1$. Indeed the function (\ref{eq:Fz})
\begin{equation}\label{eq:FzHurwitz}
F(s)=k_I+ k_P s+ k_D s^2 + \frac{B(s)}{A(s)}
\end{equation}
on the imaginary axis $s=j\omega$ decouples into two equations of the form (\ref{eq:singFreqsLines}) and (\ref{eq:singFreqs}), similar (but not equal) to equations (\ref{eq:basicIdea}). Note that the rank-condition (\ref{eq:rankConditionHurwitz}) applies also on any line parallel $\sigma=\sigma_0$ to $s=j\omega$. Thus all derivations hold also for $\partial\Gamma~=\{\sigma_0+j\omega:~\omega\in \mathbb{R}\}$, see \cite{bajc:auto:2006}.

\subsection{Schur-stability}
Consider the Schur-circle
$\partial\Gamma_1~=\{e^{j\alpha}:~\alpha\in [-\pi,~\pi]\}$. It can be
easily checked that the rank-condition
(\ref{eq:RangebdingungFuerSingulaereGammaGebiete}) is satisfied
on $\partial\Gamma_1$ for
\begin{equation}
\label{eq:SchurQEquation} Q=(1+z^2) r_1+z r_2+ r_3
\end{equation}
and that the matrix $T$, as defined in (\ref{eq:Tmatrix}), reads
\begin{eqnarray}\label{eq:Tcircle} {T} = \left[
\begin{array}{ccc}
1 & 0 & 1\\0 & 1 & 0\\1 & 0 & 0
\end{array}
\right].
\end{eqnarray}
Following (\ref{eq:decouplingFuntion}), the trivial decoupling functions of (\ref{eq:SchurQEquation}) on the Schur-circle $\Gamma_1$ are
\begin{equation}\label{eq:decouplingFunctionScur}
E_\Gamma(z) = 1+z^2~~\textrm{or}~~E_\Gamma(z) = z.
\end{equation}
For $E_\Gamma(z) = z$, the imaginary part of the function on $\Gamma_1$
\begin{equation}\label{eq:FzSchur}
F(z)=\frac{1+z^2}{z} r_1+r_2 + \frac{B(z)}{z A(z)}r_3
\end{equation}
is of the form (\ref{eq:Iq1}), since imaginary part of $\frac{1+z^2}{z}$ is null on $\partial\Gamma_1$.

\subsection{$\Gamma$-stability}
\begin{figure}[t]
\begin{center}
\includegraphics[width=7cm]{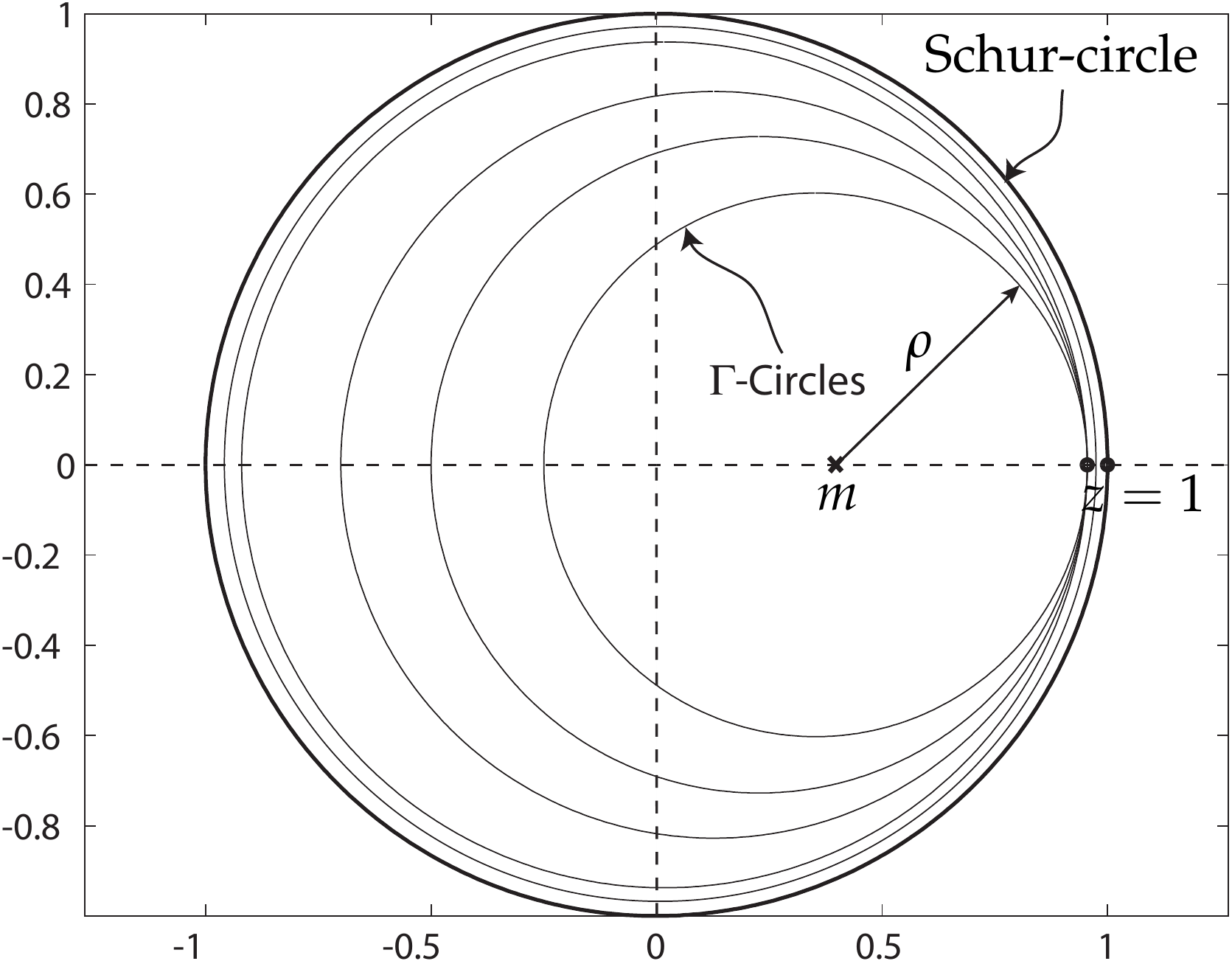}
\caption{Schur- and $\Gamma-$circles}\label{fig:circles}
\end{center}
\end{figure}

Consider a $\Gamma-$circle with center on real axis $\partial\Gamma~=\{m+\rho
e^{j\alpha}$, $\alpha\in [-\pi,~\pi]\}$, Fig.~\ref{fig:circles}.
Now
\begin{equation}
\label{eq:CirclesQEquation} Q=(\rho^2-m^2+z^2) r_1+(z-m) r_2+ r_3.
\end{equation}
For $\Gamma-$circles with center at $\tau=m$ and radius $\rho$ a
transformation matrix from $c-$ to $r-$parameter space is
\begin{eqnarray}\label{eq:12} {T} = \left[
\begin{array}{ccc}
\rho^2 - m^2 & -m & 1\\0 & 1 & 0\\1 & 0 & 0
\end{array}
\right].
\end{eqnarray}
Furhtermore
\begin{equation}\label{eq:decouplingFunctionCircle}
E_\Gamma(z) = z-m,
\end{equation}
and
\begin{equation}\label{eq:FzCircle}
F(z)=\frac{\rho^2-m^2+z^2}{z-m} r_1+r_2 + \frac{B(z)}{(z-m) A(z)}r_3.
\end{equation}

\section{Stable convex polygons}
\label{sec:stablePolygons}
This section recalls briefly a solution to problem [P2] as formulated in Section~\ref{sec:veryBasicIdea}.
For details the reader is referred to \cite{bajc:med2001} and \cite{bajc:auto:2006}. The
algorithm is motivated by the concept of inner polygons, which claim a necessary condition
for stability: A polygon $\Pi$ is said to be an inner polygon if entering the
polygon in ($r_1,r_2$)- i.e. ($k_I,k_D$)-plane, causes an eigenvalue-pair to enter
the $\Gamma-$region at the corresponding singular frequency.

\subsection{Discrete-time domain}
\begin{figure}[h]
\begin{center}
{\includegraphics[scale=0.7]{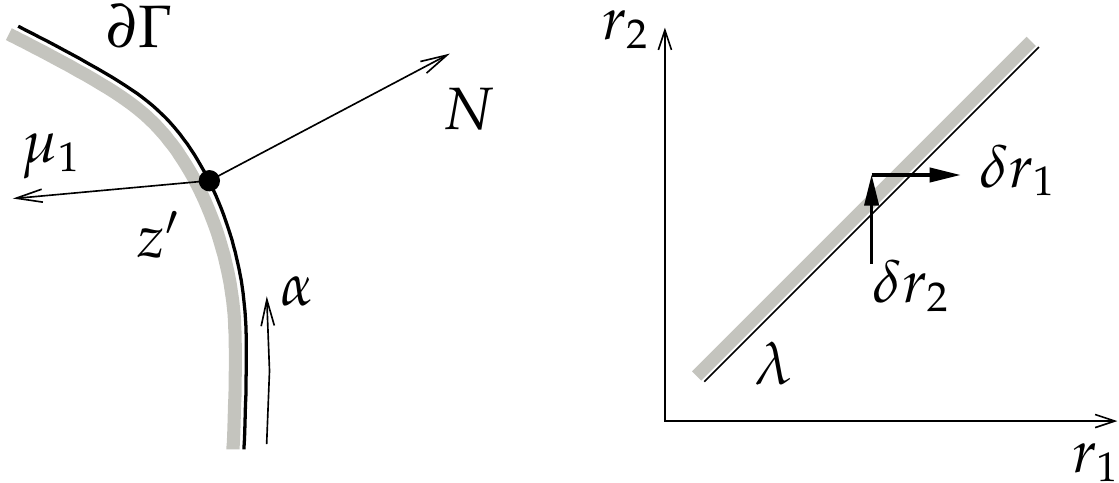}} \caption{Definition
of $e_1$ and $e_2$: the motion of eigenvalues in the vicinity of a
singular frequency $z'$. The shaded part refers to
the stable side of $\Gamma$ and the normal $N$ points outside the
$\Gamma$-region.} \label{fig:e1e2defs}
\end{center}
\end{figure}

Let $\{z'\}$ be the set of singular frequencies on $\partial\Gamma$ determined by the equation
(\ref{eq:singFreqs}) for a fixed $r_3$, and let $\{\lambda=\lambda(z')\}$ be the corresponding straight lines determined by (\ref{eq:singFreqsLines}), see Fig.~\ref{fig:e1e2defs}. In order to automate the detection of an inner polygon each boundary line $\lambda$, will be assigned a "transition" $e$: it is negative if the transition $[\delta{}r_1,\delta{}r_2]$ (see Fig.~\ref{fig:e1e2defs}) over the singular line causes an eigenvalue to become stable, i.e. enter the $\Gamma$-region, otherwise it is positive. Let $e_1$ correspond to $\delta{}r_1>0,~\delta{}r_2=0$ and $e_2$ to $\delta{}r_2>0,~ \delta{}r_1=0$.

To compute the functions $e_1$ and $e_2$, define the normal vector
$\bmf{N}$ on $\partial\Gamma$ at the singular frequency $z'=\tau(\alpha')\pm j\eta(\alpha')$ by its complex associate as
\begin{equation}
    \bmf{N}=\left({\partial F}/{\partial\tau}+j {\partial F}/{\partial\eta}
      \right)_{\normalsize z'}.
\end{equation}
For tracking the motion of an eigenvalue due to small variations in $r_1$ and $r_2$ introduce the vectors
\begin{equation}\label{eq:eid=n*muid}
{\mu_1}=\frac{\displaystyle\textrm{d}{z}}{\displaystyle\textrm{d}{r_1}}
=\frac{\displaystyle\textrm{d}\tau}{\displaystyle\textrm{d} r_1}+j\frac{\displaystyle\textrm{d}\eta}{\displaystyle\textrm{d}r_1}
~~\textrm{and}~~\mu_2=\frac{\displaystyle\textrm{d}{z}}{\displaystyle\textrm{d}{r_2}}.
\end{equation}
Assuming ${\partial{p}}/{\partial{z}} \neq 0$\footnote{For situations with ${\partial{p}}/{\partial{z}} = 0$, refer to \cite{bajc:auto:2006}.}, it can be easily shown that
\begin{equation}\label{eq:mu1_complex}
{\mu_1}=-{\displaystyle\frac{\partial{p}}{\partial
{r_1}}}/{\displaystyle\frac{\partial{p}}{\partial {z}}}.
\end{equation}
Now transitions functions can be computed by
\begin{equation}\label{eq:eid=n*muid}
e_{1/2}=\mathrm{Re}\left(\mu_{1/2}^{\ast}N\right)_{\normalsize z'}.
\end{equation}
Using this information, an algorithm for the detection of the
\emph{inner polygons} (those with maximal number of $\Gamma$-stable eigenvalues) is
developed in \cite{bajc:med2001}. Such polygons are the only stability candidates, that can be proved by checking any point therein.

\begin{figure}[h]
\begin{center}
{\includegraphics[width=7.5cm]{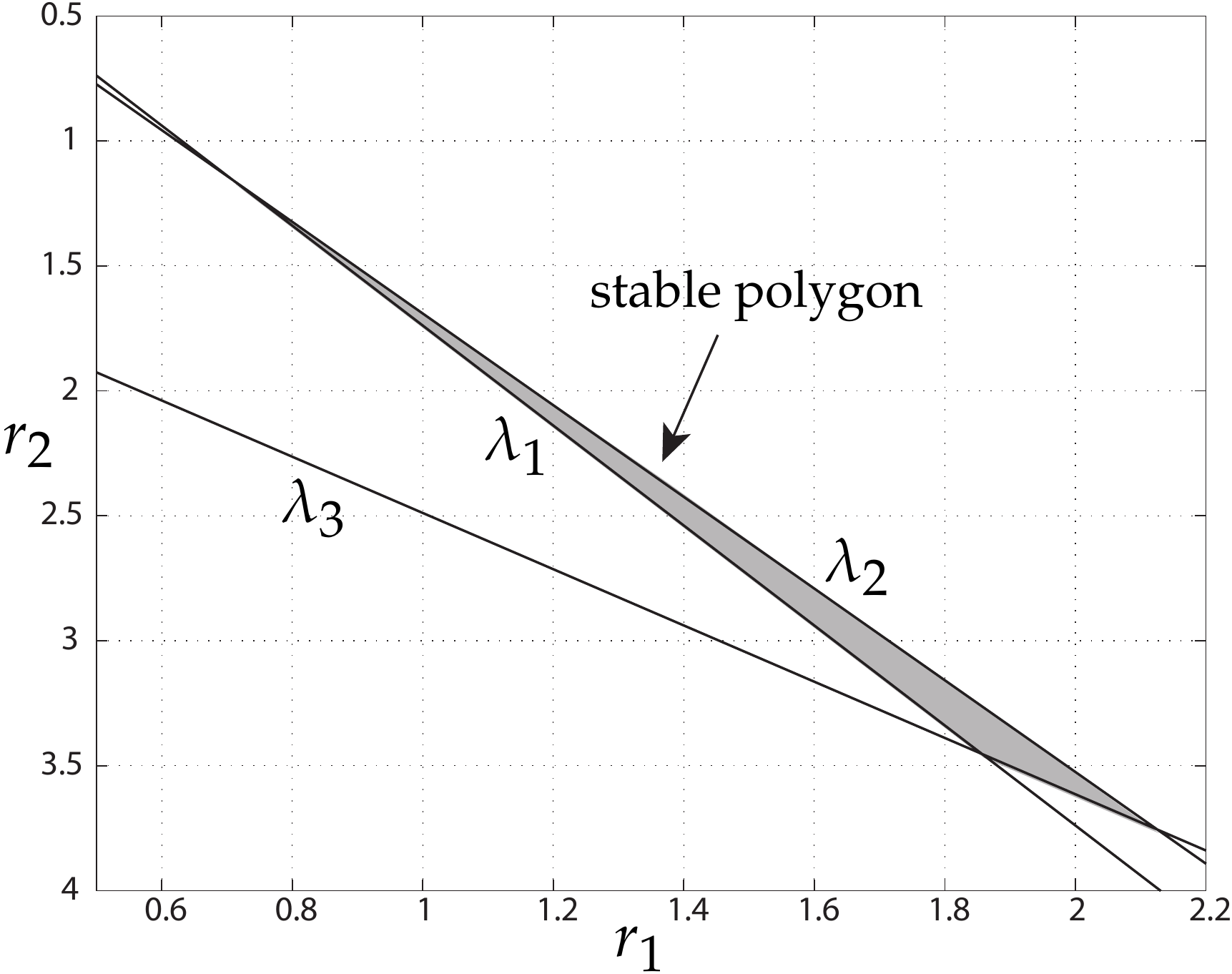}}
\caption{The stable polygon for $r_3=-0.26118$, Example 1}
\label{fig:astabpolyg1}
\end{center}
\end{figure}

{\small
\emph{Example~1:}~Consider the discrete-time model of the plant
\begin{equation}\label{eq:GsDigit} G=10^{-6}\frac{4.165 z^3 +
45.77 z^2 + 45.77 z + 4.165}{z^4 - 3.985z^3 + 5.97z^2 - 3.985z +
1}
\end{equation}
and a three-term stabilizer
\begin{equation}
\label{eq:threeTerm} C(z)=10^4\frac{(z^2 - 1.541 z
+0.5992)(c_1+c_2 z+c_3 z^2)}{z(z+0.4047)(z+0.2162)(z-0.4934)},
\end{equation}
whose parameters $c_1, c_2, c_3$ are to be synthesized. The
synthesis is done in $(r_1,r_2,r_3)$-parameter space. Therefore
the transformation (\ref{eq:Tcircle}) can be used. Then
\begin{eqnarray*}\label{eq:Azexmaple}
A(z)&=& {z}^{5}+9.44z^4-5.34 z^3-9.34z^2+5.04 z+0.59\\
\label{eq:Bzexmaple}
B(z)&=&0.19 z^8-0.73 z^7+z^6-0.45 z^5-0.12 z^4+\cdots\nonumber\\
{} &{} & 0.14 z^3-0.009z^2-0.008z.
\end{eqnarray*}
For $r_3=-0.26118$, the singular frequencies lying on the Schur-circle are computed to
be $z'_1=1, z'_{2}=0.9172\pm j 0.3983, z'_{3}=0.5628\pm j
0.8266, z'_{4}=-1$. The resulting stable polygon is shown in
Fig.~\ref{fig:astabpolyg1}. It is enclosed by the straight lines
frequencies $\lambda_1,\lambda_2$ and $\lambda_3$, corresponding
to $z'_1, z'_2 ~\textrm{and}~ z'_3$.}

\subsection{Hurwitz conditions}
Let $e_I$ correspond to $\delta{}k_I>0,~ \delta{}k_D=0$ and $e_D$
to $\delta{}k_D>0,~ \delta{}k_I=0$. It can be shown that (\ref{eq:eid=n*muid}) yields
\begin{equation}
    \label{eq:cros}
    e_{I/D}=\Bigg\vert\frac{\partial(H,G)}{\partial(\omega,k_{I/D})}\Bigg\vert_{\normalsize
    \omega'}.
\end{equation}
Expressions for $e_{I/D}$ do not depend on where a boundary line is crossed at. Indeed, check that
(\ref{eq:cros}) is equivalent to
\begin{equation}
    \label{eq:signsEquation2}
    e_{I/D}=\frac{\partial k_P}{\partial\omega}\Bigg\vert\frac{\partial(H,G)}{\partial(k_P,k_{I/D})}\Bigg\vert_
    {\normalsize\omega'},
\end{equation}
where $k_P=k_P(\omega)$ stands for the $k_P$-plot. Since $\omega'$ is a singular frequency, the determinant in the latter equation is shown to be independent on $k_I$ and $k_D$. Furthermore, the following holds
\begin{equation}
    \label{eq:signsEquation3}
    \mathrm{sign}~e_{I}=-\mathrm{sign}~\frac{\partial
    k_P}{\partial\omega}\Bigg\vert_{\omega'},~~~~~~
    \mathrm{sign}~e_{D}= \mathrm{sign}~\frac{\partial
    k_P}{\partial\omega}\Bigg\vert_{\omega'}.
\end{equation}
These expressions prove again that the transitions $e_I$ and $e_D$ are
independent on parameters $ k_I$ and $ k_D$, and of opposite sign.
Their sign is determined by the slope of the $k_P$-plot at the
corresponding singular frequency, see Fig.~\ref{fig:kpplot}.

\begin{figure}[h]
    \centering
    \includegraphics[width=7.5cm]{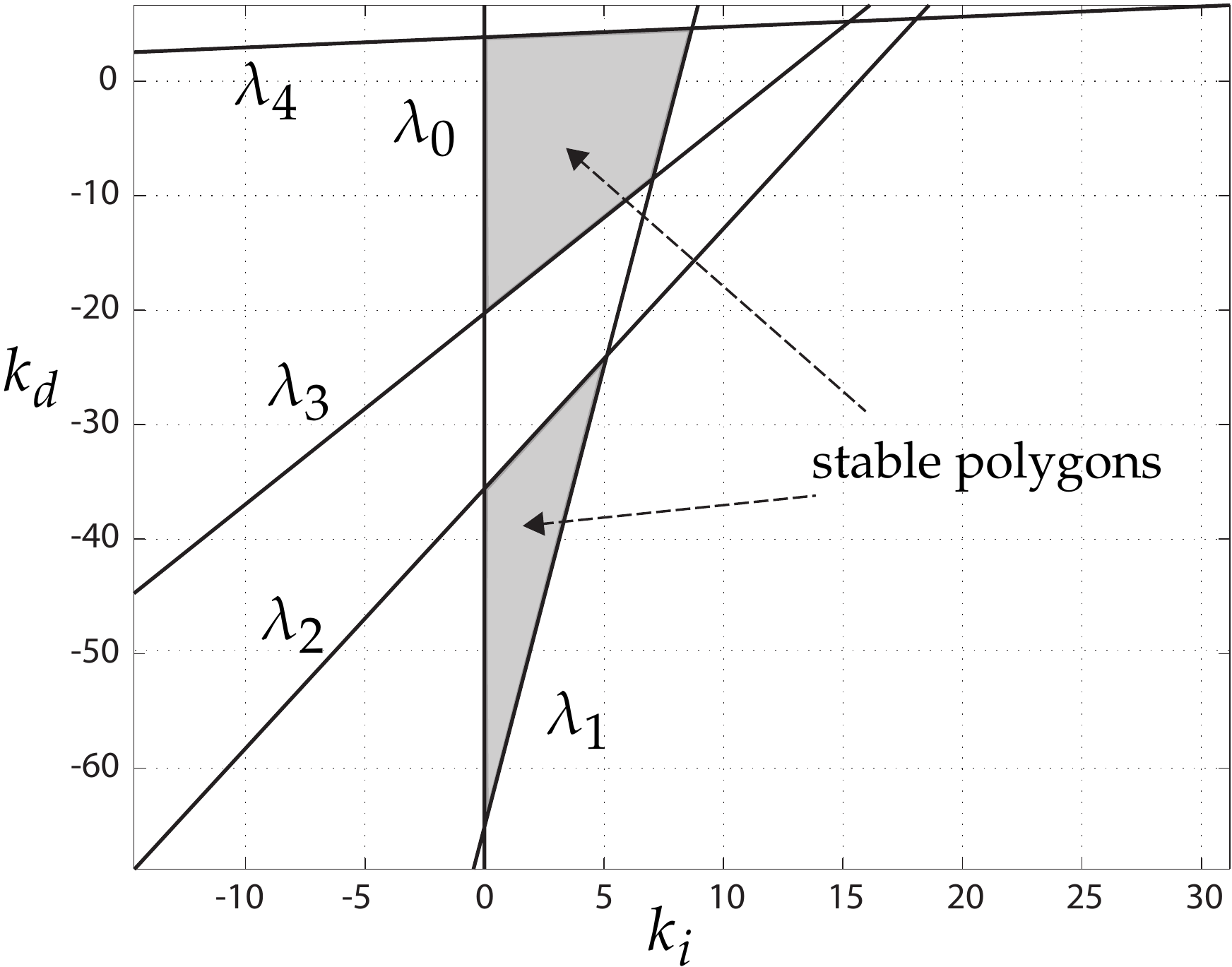}
    \caption{Stable polygons for $k_P=-2$, Example~2}
    \label{fig:singLinesExample1}
\end{figure}

{\small
\emph{Example~2:}~Consider the polynomial
(\ref{eq:basicCharacteristicPolynomial}) with
\begin{eqnarray*}\label{eq:cpExample1}
A(s)&=& -0.5 s^4-7 s^3-2 s+1\\
\label{eq:cpExample11}
B(s)&=&{s}^{7}+11\,{s}^{6}+46\,{s}^{5}+95\,{s}^{4}+109\,{s}^{3}+74\,s^2+24
s.
\end{eqnarray*}
The singular frequencies for $k_P=-2$ are computed from its $k_P$-plot (see Fig.~\ref{fig:kpwGraphExample1}): $s'_0=0,~s'_1=\pm j 0.3530,~s'_2=\pm j 0.6638,~s'_3=\pm j
0.7742,~s'_4=\pm j 3.3473.$ Fig.~\ref{fig:singLinesExample1} depicts the corresponding straight lines and the stable polygons. Note that the stable polygons need not to be connected.}

\section{$k_P$-Problem} \label{sec:kp-Porblem}
\subsection{Hurwitz-stability} \label{sec:kp-Porblem:Hurwitz}
This section focuses on the problem [P1], as defined in Section~\ref{sec:veryBasicIdea}: a rule is sought to discriminate $k_P-$intervals with stable PID controllers. Intuitively, it must be tightly related
to the $k_P$-plot. Indeed, it is clear that at maxima and minima of the $k_P-$plot - compare Fig.~\ref{fig:kpplot} and \ref{fig:kikdplane} - convex polygons close at an edge due to the overlapping of two straight boundary lines. As $k_P$-intervals with maximal number of singular frequencies are most likely to host stable polygons. The following result renders this idea precisely.

{\it
\begin{thm}\label{thm:r2GriddingIntervalsHurwitz}
Consider the polynomial (\ref{eq:basicCharacteristicPolynomial}).
Assume that polynomial $A(s)$ has no zeros on the imaginary
axis and let\\
\begin{tabular}{ll}
$N$: & order of the polynomial (\ref{eq:basicCharacteristicPolynomial})\\
$M$: & order of the polynomial $A(s)$\\
$P$: & number of RHP zeros of $A(s)$\\
$Z$: & number of singular frequencies in the interval\\
{} & {} $0 < \omega < +\infty$.\\
\end{tabular}

A necessary condition for stability of
(\ref{eq:basicCharacteristicPolynomial}) is
\begin{equation}
    \label{eq:kPConditionHurwitz}
    Z\geq \frac{E(N-M+2P-1)}{2}.
\end{equation}
\end{thm}}
Here, the function $E:\mathbb{N}\mapsto\mathbb{N}$ selects the nearest smaller even number. The proof of the theorem can be found in \cite{bajc:auto:2006}. It is important to observe, that we exclude the zero singular frequency from $Z$.

Using this criterion one can directly read from the $k_P$-plot (Fig.~\ref{fig:kpplot}) the
$k_P-$interval(s) with (potentially) host stable polygons. However, in some cases a polygon may close
when three boundary lines intersect at a single point in ($k_P, k_I, k_D$)-parameter space. This situation destroys the sufficiency of the condition (see Lemma~\ref{thm:stabPeaks}) is not sensed by the above criterion and will be discussed in Section~\ref{sec:stabilitypeaks}.

The following is an extension of Theorem~\ref{thm:r2GriddingIntervalsHurwitz} to the cases when
$A(s)$ possesses zeros on the imaginary axis.

{\it
\begin{thm}\label{thm:r2GriddingIntervalsHurwitz2}
Suppose $A(s)$ has $J$ zeros on the imaginary axis. Then, for stability of the
polynomial (\ref{eq:basicCharacteristicPolynomial}) $Z$, singular frequencies
are required within the interval $\omega\in(0,+\infty)$, where

(a)~~if $s=0$ is not a zero of $A(s)$\\
\begin{equation}
    \label{eq:kPConditionHurwitz_1}
    Z \geq \frac{E(N-M+2P-J-1)}{2}
\end{equation}

(b)~~if $s=0$ is a zero of order $J_0$ of $A(s)$\\
\begin{equation}
    \label{eq:kPConditionHurwitz_2}
    Z \geq \frac{E(N-M+2P-J-1)-E(J_0)}{2}.
\end{equation}
\end{thm}}

\begin{figure}[t]
    \centering
    \includegraphics[width=8cm]{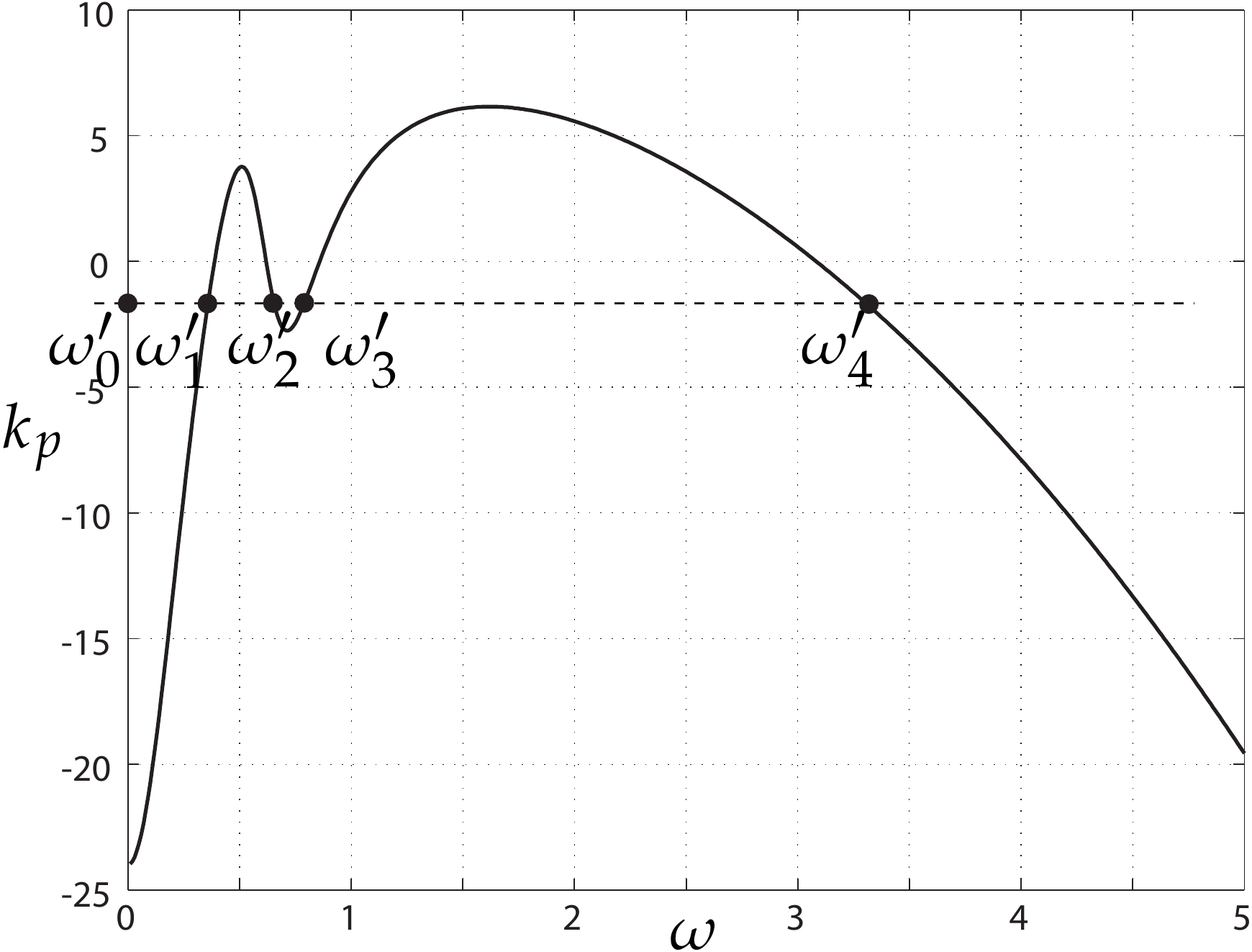}
    \caption{The $k_P$-plot, Example~2}
    \label{fig:kpwGraphExample1}
\end{figure}
\begin{figure}[h]
    \centering
    \includegraphics[width=8.25cm]{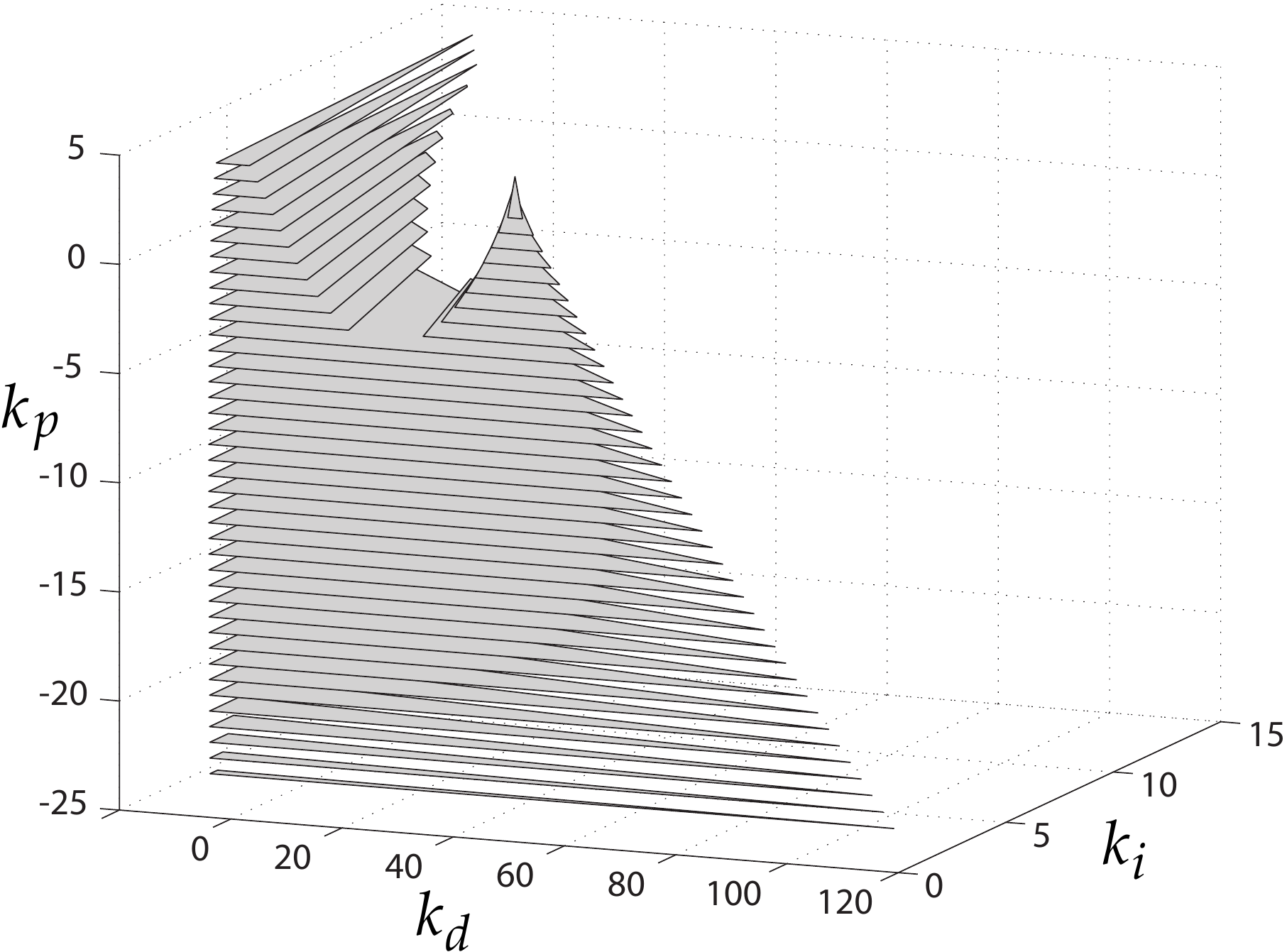}
    \caption{ The region of PID stabilizers, Example~2}
    \label{fig:3DsingLinesExample1}
\end{figure}

{\small
\emph{Example~2: (cont).}~ The $k_P$-plot is depicted in Fig.~\ref{fig:kpwGraphExample1}. Notice that it is very easy to read the number of available singular frequencies from Fig.~\ref{fig:kpwGraphExample1} for a given $k_P$. For the polynomial $A(s)$ we have $N=7,~M=4,~P=1,~J=0,~J_0=0$. According to
Theorem~\ref{thm:r2GriddingIntervalsHurwitz} for stability at least $Z \geq {E(N-M+2P-J-1)}/{2}=2$ singular frequencies are required for $\omega > 0$. By checking the Fig.~\ref{fig:kpwGraphExample1} it is obviously that this condition is fulfilled for $-24<k_P<6.1565$. More precisely, for $-24 <  k_P < -2.7614$ and $3.7664 < k_P < 6.1565$, two non-zero singular frequencies exist, and for $-2.7614 < k_P < 3.7664$ four ones. Finally, by gridding $k_P$ within these intervals, stable polygonal slice can be computed. The result is shown in Fig.~\ref{fig:3DsingLinesExample1}.}

{\small{\emph{Example~3:~Separated stable $k_P-$intervals.~} Consider the
polynomial (\ref{eq:basicCharacteristicPolynomial}) with
\begin{eqnarray*}
A(s)&=& s^3+3s^2+9\\
B(s)&=&s^5+2 s^4+3 s^3+7 s^2+14 s.
\end{eqnarray*}
It can be directly read that $N=5,~M=3$ and $P=2$. According to Theorem~\ref{thm:r2GriddingIntervalsHurwitz} for
stability at least $2$ positive singular frequencies are required. Now consider
Fig.~\ref{fig:kpw2Intervals}, where the $k_P$-plot for $\omega\geq
0$ is depicted. Two $k_P$-intervals of interest are
directly recognized, namely $-1.8708 <  k_P < -1.5556 $ with 3 positive singular~frequencies,
and $0.3157 < k_P < 0.5333$ with 4 ones. For other $k_P$s no stable polygons exist.}

\begin{figure}[h]
    \centering
    \includegraphics[width=8cm]{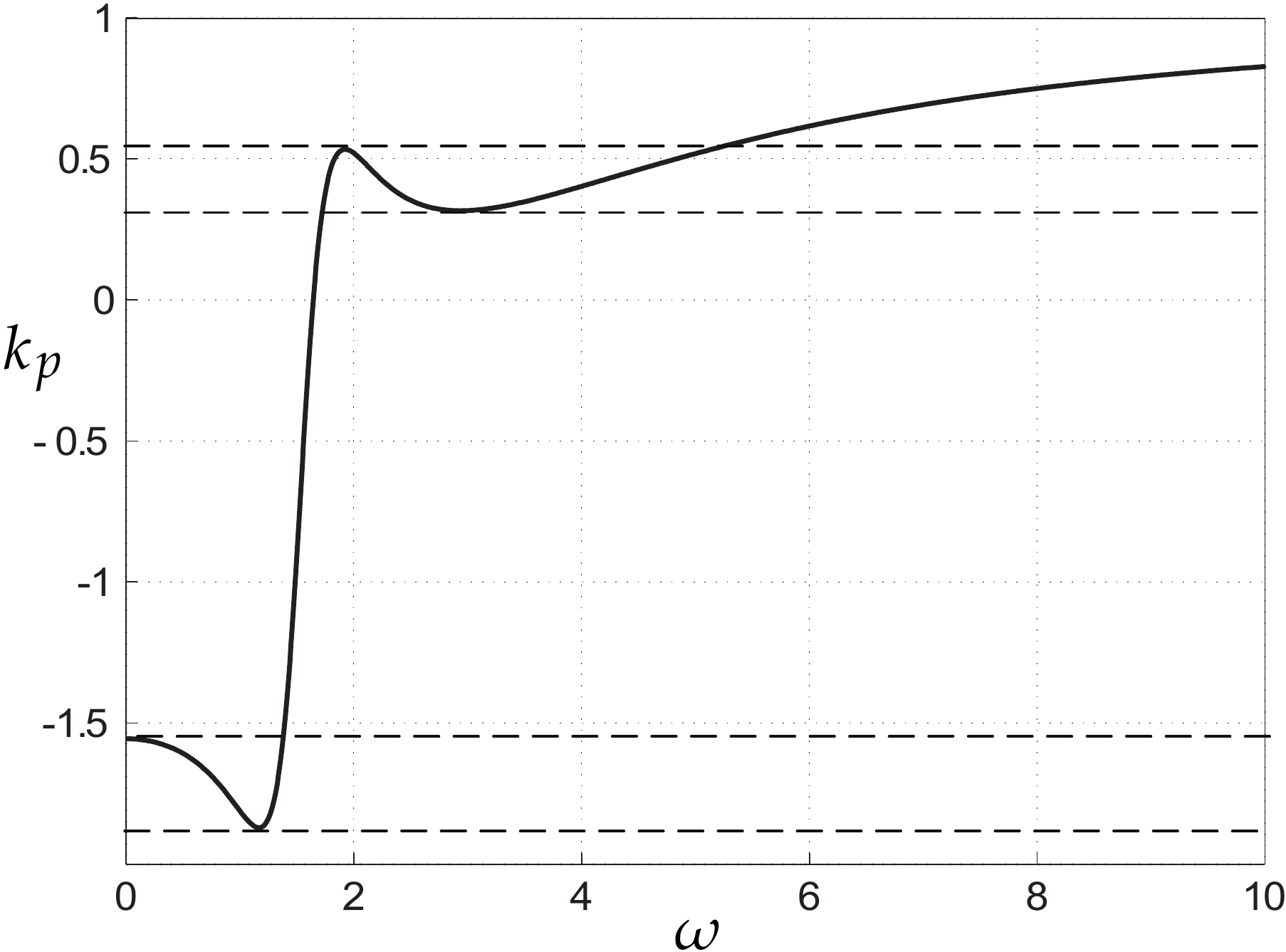}
    \caption{The $k_P$-plot, Example~3}
        \label{fig:kpw2Intervals}
\end{figure}

{\small
\emph{Example~4: Missing stability.~} Let
\begin{eqnarray*}
A(s)&=& 1\\
B(s)&=&s^5+s^4-3s^3-s^2+2 s.
\end{eqnarray*}
Theorem~\ref{thm:r2GriddingIntervalsHurwitz} requires at least $2$
singular frequencies in $\omega>0$, however for $-2>k_P$, $1$ singular
frequency exists, otherwise none. Thus, polynomial
(\ref{eq:basicCharacteristicPolynomial}) is always unstable, no matter
what $k_P,~k_I,~k_D$.}}

\subsection{Schur-stability} \label{sec:kp-Porblem:Schur}
Without loss of generality, we consider just the Schur-circle. The generalizations for other
$\Gamma-$circles are straightforward.
{\it
\begin{thm}\label{thm:r1GrddingIntervals}
Consider the characteristic polynomial (\ref{eq:basicEquation})
and the Schur-circle $\Gamma_1$. Let\\
\begin{tabular}{ll}
$N$: & order of the polynomial (\ref{eq:basicEquation})\\
$R$: & number of zeros of $A(z)E_\Gamma(z)$ lying inside $\partial\Gamma_1$\\
$J$: & number of zeros $\neq\pm 1$ of $A(z)E_\Gamma(z)$ lying on $\partial\Gamma_1$\\
$J_+$: & order of the zero $+1$ of $A(z)E_\Gamma(z)$ \\
$J_-$: & order of the zero $-1$ of $A(z)E_\Gamma(z)$ \\
$Z$: & number of singular frequencies in the interval \\
{}& $0 < \alpha < +\pi$.
\end{tabular}

A necessary condition for stability of (\ref{eq:basicEquation}) is
\begin{equation}
    \label{eq:kPCondition}
    Z \geq N-R-\frac{J+E(J_+)+E(J_-)+2}{2}.
\end{equation}
\end{thm}
}

\begin{figure}[h]
\begin{center}
{\includegraphics[width=8cm]{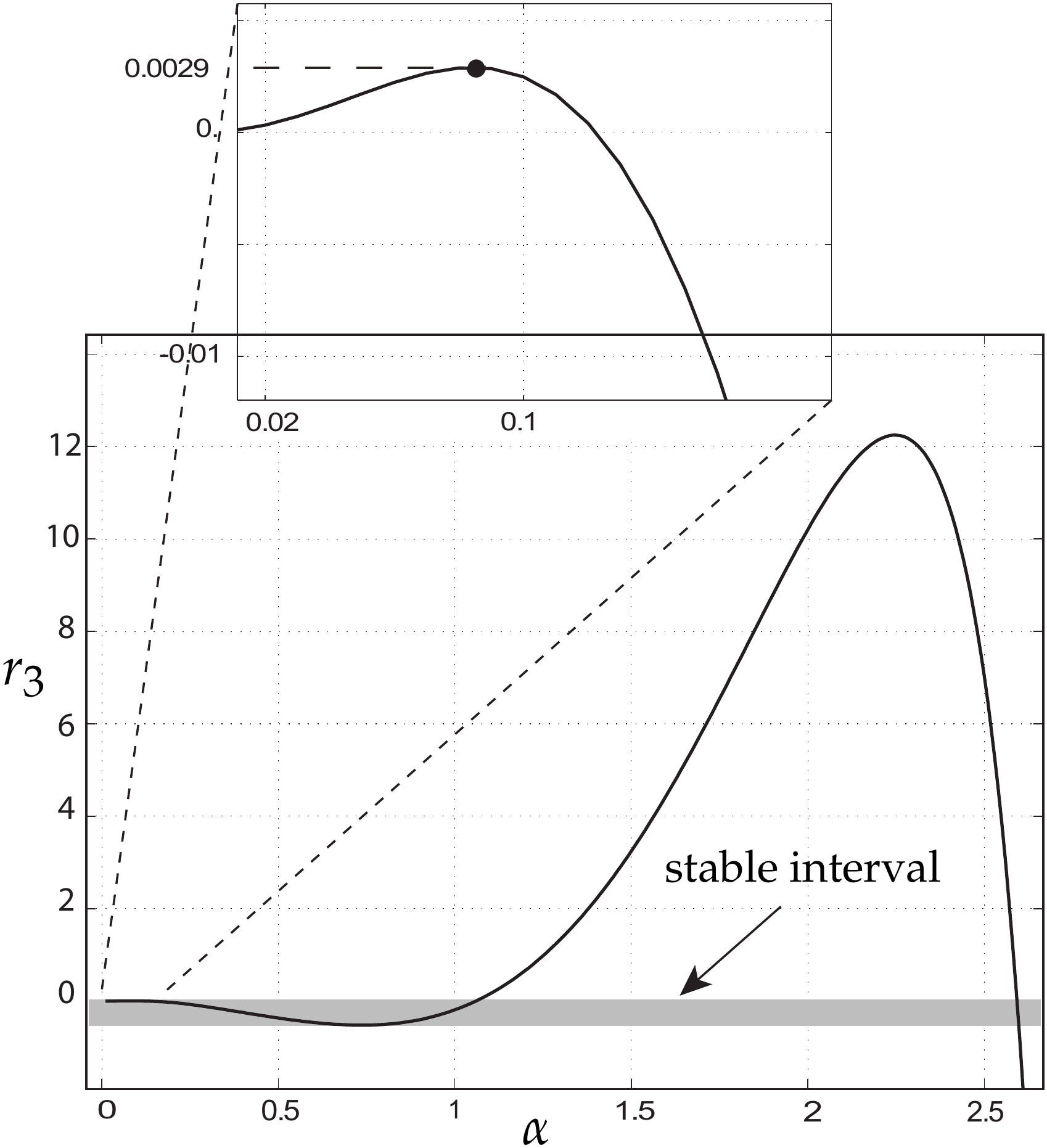}} \caption{The
$k_P$-plot, Example~1}
\label{fig:kpwexample}
\end{center}
\end{figure}

{\small \emph{Example~1: (cont).}~It can be checked
that $A(z)$ has three zeros inside the Schur-circle, one
zero at $z=-1$ and one zero outside the Schur-circle. Thus if the
decoupling function $E_\Gamma(z)=z$ is used, it follows that $N=8,~R=3+1=4, ~J=1,~J_+=0$, and $J_-=1$.
Hence, for stability, $Z\geq 3$ singular frequencies are required
in the interval $0<\alpha<+\pi$. In order to discriminate stable
$r_3$ intervals check the $k_P$-plot in Fig.~\ref{fig:kpwexample}. The
stable interval is indicated by the grayed strip in
Fig.~\ref{fig:kpwexample} $-0.52236 < r_3 <0.00290$.
Notice that the zoomed plot in Fig.~\ref{fig:kpwexample}
reveals that for $0< r_3 <0.00290$ four additional singular
frequencies appear.

On the other side if the decoupling function $E_\Gamma(z)=1+z^2$
is used, then $N=8,~R=3,~J=3,~J_+=0$, and $J_-=1$ i.e. again for stability $Z\geq 3$ singular frequencies are required in the interval $0<\alpha<+\pi$.}

%
%

{\small
\emph{Example~5:~PID control.}~Consider PID control of the plant (\ref{eq:GsDigit}) now using
the control law (\ref{eq:PID}). It can be shown that in this case
\begin{eqnarray}\label{eq:A1B1}
A(z)&=&{z}^{3}+ 10.98\,{z}^{2}+10.98\,z+ 1\\
B(z)&=&0.1 z^6-0.5 z^5+z^4-z^3+0.5 z^2-0.1 z.
\end{eqnarray}

By using $E_\Gamma(z)=z$, it is easily checked that
$N=6,~R=1+1=2, ~J=0,~J_+=0$ and $J_-=1$. Hence, $Z\geq 3$ singular
frequencies within $0<\alpha<+\pi$ are required. However, the maximal number of singular frequencies within $0<\alpha<+\pi$ is $2$, so no PID controller can stabilize the
plant (\ref{eq:GsDigit}).}

\subsection{Stability peaks}\label{sec:stabilitypeaks}

{\small{
\emph{Example~6:~Stability peaks.} Consider polynomials
\begin{eqnarray*}\label{eq:Example4}
A(s)&=& 1890\,{s}^{2}+658\,s+215\\
B(s)&=&{s}^{8}+{\frac {1032}{25}}\,{s}^{7} +{\frac
{6175327}{10000}}\,{s}^{6}+{\frac
{98620159}{25000}}\,{s}^{5}+ \\
{}& {} &{\frac {92785263}{10000}}\,{s}^{4}+{\frac
{97588159}{25000}}\,{s^3}+{\frac {5413746}{625}}\,{s}^{2}.
\end{eqnarray*}
According to definitions in Theorem~\ref{thm:r2GriddingIntervalsHurwitz}, $N=8,~M=2,~P=0$.
Therefore for stabilizability $Z\geq{E(N-M+2P-1)}/{2}=3$ singular
frequencies are required in the interval $0 < \omega<+\infty$. It
can be shown that for both, $k_P=-9$ and $k_P=-10$, $Z=3$
positive singular frequencies are available, which satisfy
the stabilizability condition. But, for $k_P=-9$ a stable polygon
exists, and for $k_P=-10$ it does not. In other words, for some $k_P$ in
between, the stable polygons must close in a vertex (and not an edge). To handle this
situation one has to detect stability peaks within the
intervals provided by Theorem~\ref{thm:r2GriddingIntervalsHurwitz}.}}

Note that at a peak, the three-term polynomial (\ref{eq:basicCharacteristicPolynomial}) must have
at least three different eigenvalues on the imaginary axis, that is
\begin{equation}\label{eq:peaks}
A(s)(k_I+k_Ps+k_D s^2)+B(s)=R(s)\prod_{i=1,2,3}(s^2+\omega'^2_i),
\end{equation}
where $R(s)$ is a remainder polynomial, which has to be stable,
otherwise the peak is irrelevant. This is a nonlinear set of $N$
equations with $N$ unknowns. The left-hand side of the equation
provides the three unknowns $k_I,k_P,k_D$ and the right-hand side
the rest $N-3$ ones, including three singular frequencies
$\omega'_i,~i=1,2,3$ and $N-6$ coefficients of the polynomial
$R(s)$. Hence by elimination of the latter $N-3$ variables, a
system of three nonlinear equations with the three $k_I,k_P,k_D$
variables results. Its solution provides the required peaks. In
general, finitely many stability peaks exist.


{\small
\emph{Example~6: (cont).}~It can be shown that for the three-term polynomial defined by (\ref{eq:Example4}) a stability peak appears at $k_P\approx-9.0023$. The three straight lines corresponding to $\omega'_1\approx 0.2581$, $\omega'_2\approx 0.44261$
and $\omega'_3\approx  9.7621$, intersect at
$k_D\approx 21.4958, k_I\approx 3.0195$. A stability peak appears also in Example~7, see Fig.~\ref{fig:det6}.}

{\it
\begin{lem}\label{thm:stabPeaks}
The condition in Theorem~\ref{thm:r2GriddingIntervalsHurwitz} is also sufficient for $N\leq 6$. Theorem~\ref{thm:r2GriddingIntervalsHurwitz} provides necessary and sufficient conditions for any PID feedback loop with plants of 1st, 2nd and 3rd order.
\end{lem}}

\section{PID for time-delay systems}\label{sec:TimeDelaySystems}
In this section we extend the PID control theory to systems with time-delay. Such feedback loops always result in quasi-polynomials of the form
\begin{eqnarray}
    \label{eq:basicEquationTD}
    p =A(s)( k_I+k_P s +k_D s^2)+ B(s)e^{L s},
\end{eqnarray}
where $L>0$ is the time-delay. Fundamental stability conditions
regarding quasi-polynomials are provided in \cite{pon:55}. E.g. a
simple necessary condition is the existence of principal term,
i.e. $e^{Ls}$ in (\ref{eq:basicEquationTD}) must be multiplied by
the highest power in $s$. Thus, in this section we assume
$n\geq m+2$, i.e. the quasi-polynomials of the \emph{retarded} ($n>m+2$)
and \emph{neutral} type ($n=m+2$) are considered only.

It is easy to check that decoupling conditions hold for the quasi-polynomial
(\ref{eq:basicEquationTD}), too. With definitions
\begin{equation}
    \label{eq:amp_phase_exp}
    \alpha(\omega)=\sqrt {{\frac {{{R_{B}}}^{2}+
    {{I_{B}}}^{2}}{{{{R}_{A}}}^{2}+{{I_{A}}}^{2}}}},~~
    \tan\phi(\omega) = {{\frac {{R_{A}}\,{I_{B}}-{I_{A}}\,{R_{B}}}{{R_{A}}\,{R_{B}}+
    {I_{A}}\,{I_{B}}}}},
\end{equation}
equations (\ref{eq:singFreqsLines}) and (\ref{eq:singFreqs}) take the form
\begin{eqnarray}
\label{eq:singularLineGeneratorTD}
     k_I -\omega^2  k_D & = & \alpha(\omega) \cos(\omega L+\phi(\omega)),\\
\label{eq:singularFrequencyGeneratorTD}
    \omega k_P &=& {\alpha(\omega)}\sin(\omega  L+\phi(\omega)).
\end{eqnarray}
Obviously $k_P$-plot is now a sinusoidal function and we have to deal with infinitely many singular frequencies.

\subsection{High-frequency behavior}
\label{sec:HighfrequencyBehaviour}
Consider equation (\ref{eq:amp_phase_exp}). Given that $n\geq m+ 2$ for high singular
frequencies, i.e. as $\omega\to\infty$
\begin{equation}
    \label{eq:alpha_infty}
    \alpha(\omega)\sim\omega^{n-m}
\end{equation}
and
\begin{equation}
    \label{eq:phase_infty}
    \tan \phi(\omega)\to
    \left\{
    \begin{array}{l}
    \pm\infty \\
    0
    \end{array}
    \right.
    ~~\Rightarrow~~
    \phi(\omega)\to
    \left\{
    \begin{array}{l}
    \pm\frac{\pi}{2}\\
    0~~\textrm{or}~~\pi.
    \end{array}
    \right.
\end{equation}
For a fixed $k_P$-grid, equation
(\ref{eq:singularFrequencyGeneratorTD}) implies
\begin{equation}
    \label{eq:sf_infty}
    \sin(\omega  L +\phi(\omega)) \to
    \frac{1}{ k_P}{\omega^{1-n+m}} \to 0,
\end{equation}
that is, all higher singular frequencies tend to
\begin{equation}
    \label{eq:sing_freqs_infty_3}
    \omega\to
    \left\{
    \begin{array}{ll}
    k \pi/ L, & {}\\
    (k+\frac{1}{2}) \pi/ L,&  k\in \mathbb{N}, {k}\gg 1,
    \end{array}
    \right.
\end{equation}
and
\begin{equation}
    \label{eq:cos_freqs_infty}
    \cos(\omega  L +\phi(\omega)) \to \pm 1.
\end{equation}

To investigate the behavior of boundary lines for high frequencies, applying $\omega\to\infty$ to (\ref{eq:singularLineGeneratorTD}) reads
\begin{equation}
    \label{eq:irb}
     k_D = \pm{\alpha(\omega)}/{\omega^2}\big\vert_{\omega\to\infty}.
\end{equation}
For quasi-polynomials of neutral type, $n=m+2$ in (\ref{eq:alpha_infty}), and boundary lines
converge to so-called \emph{infinity root boundaries}
\begin{equation}
    \label{eq:irb1}
     k_D = \pm{b_n}/{a_m},
\end{equation}
For quasi-polynomials of retarded type, straight lines diverge. Note that infinity root boundaries
describe the situation with infinitely many eigenvalues arbitrarily close to the imaginary axis.


\subsection{Relevant frequency range}
\label{sec:theRelevantFrequencyRange}  According to (\ref{eq:signsEquation3}) the sign of the transition function $e_{I/D}$ at a singular frequency $\omega'$ is determined by the slope of the function $k_P=k_P(\omega)$ at $\omega'$. Hence, at successive singular frequencies corresponding to a fixed $k_P$,
$e_{I/D}$ takes opposite signs. This motivates the definition of the set of odd $\Omega_o$ and even $\Omega_e$ singular
frequencies
\begin{equation}
    \Omega_o=\{\omega'_1,~\omega'_3,~\omega'_5,~\cdots\},~~\Omega_e=\{\omega'_0,\omega'_2,~\omega'_4,~\cdots\}
\end{equation}
with
\begin{equation}
    0=\omega'_0<\omega'_1<\omega'_2<\omega'_3<\cdots<\infty.
\end{equation}
The transitions $e_{I/D}$ have the same sign for all even (odd) singular frequencies, which is opposite to that of odd (even) singular frequencies.

The intersection of a singular line with $k_I=0$ and fixed $k_P$
\begin{equation}
     k_D(0)=-\frac{ k_P}{\omega\tan(\omega L +\phi(\omega))}
\end{equation}
discriminates between even and odd singular lines, too. For quasi-polynomials of retarded type (no infinity root boundaries), as $\omega\to\infty$, one group of the boundary straight lines diverges toward $+\infty$, while the other towards $-\infty$. Since for high frequencies, $k_D$ and $e_D$ share the same sign, the stable region must lie between even and odd boundary lines. Thus, starting from a sufficiently large singular frequency, the boundary lines become irrelevant. The same holds for the quasi-polynomials of neutral type. However, the infinity root boundaries (\ref{eq:irb}) may impact the stable polygons: indeed these must lie between the two infinity root boundaries (\ref{eq:irb1}). As a conclusion, stable inner polygons are determined by the low frequency boundaries and infinity root boundaries (if any). For instance, it was shown that for PID control of a first order proportional term $G=K/(T s+1)e^{-Ls}$, the relevant singular frequencies are the first two ones and the two infinity root boundaries.

It is difficult to state a rule, which would precisely discriminate the relevant frequency range in the general case. However, some estimates are still thinkable. The relevant frequency range should
comprise the singular frequencies, which correspond to the minima and maxima of the stable $k_P$-intervals. Another helpful rule of thumb is to discriminate the region, where $k_P(\omega)$, Fig.~\ref{fig:kpplot}, oscillates with an almost fixed period as settled in (\ref{eq:phase_infty}).

\subsection{Stable $k_P$ intervals}
\label{sec:theStableKPIntervals} In this section the solution of
$k_P-$problem is extended to time-delay systems.

{\it
\begin{thm}\label{thm:kpGriddingIntervalsTD}
Consider the quasi-polynomial (\ref{eq:basicEquationTD}). Assume
that $A(s)$ has $J$ non-zero zeros on the imaginary axis, $P$
right-hand-side zeros and a zero of order $J_0$ at $s=0$. If (\ref{eq:basicEquationTD}) is Hurwitz-stable for a fixed
$k_P=k'_P$, then a $k \in \mathbb{N}$ exists,
such that for $l \geq k$, $l \in \mathbb{N}$, the number of
singular frequencies $Z$ in the interval $0 <
\omega < (2 l \pi+\delta)/L$ corresponding to $k_P=k'_P$ is
\begin{equation}
\label{eq:r1ConditionQuasipolynomial}
Z \geq \frac{E(4l+N-M+2P-J-1)+E(J_0)}{2},
\end{equation}
where $\delta$ is chosen such that the principal term does not
vanish at $\omega = (\pm 2l \pi+\delta)/L$.
\end{thm}}

For the proof of this theorem is refer to \cite{bajc:tds2004}.

\begin{figure}
    \centering
    \includegraphics[scale=0.5]{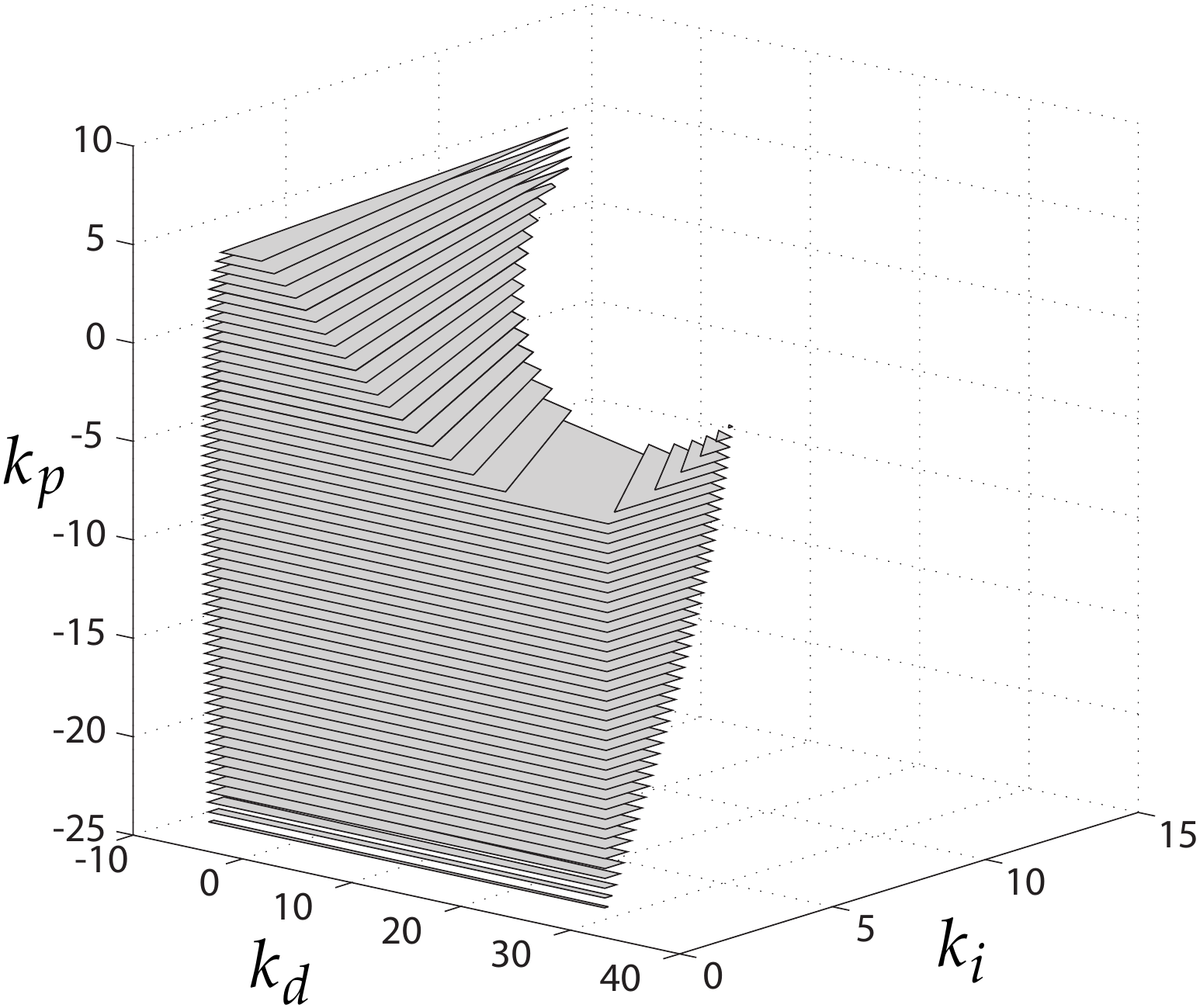}
    \caption{PID stability region, Example~7}
    \label{fig:det6}
\end{figure}

{\small
\emph{Example 7.~} Consider PID control of the plant
\begin{equation}
    \label{eq:norbertsBeispiel}
    G(s)= \frac{-s^4-7s^3-2
    s+1}{(s+1)(s+2)(s+3)(s+4)(s^2+s+1)}e^{-0.05 s}.
\end{equation}
Note that $N=7$, $M=4$ and $P=1$ (since a right-hand-sided zero of $A(s)$
exists at $s=0.3483$). According to
Theorem~\ref{thm:kpGriddingIntervalsTD}, we need to find a sufficiently
large $k$, such that within any interval $0 < \omega <40 l\pi +\delta$, with
$l\geq k$, at least $E(4 l+ 6)/2=2r+3$ singular frequencies are
available. It can be easily checked that already for $k=1$ and
$\delta=\pi$ the condition is fulfilled within $-24<k_P<6.0693$.
Fig.~\ref{fig:det6} shows the set of all PID stabilizers for the
plant (\ref{eq:norbertsBeispiel}). This example illustrates two
interesting situations: first, for $-3.7671 < k_P <4.6807$ the
stable region includes two separated polygons, and second, one of
the polygons closes at a vertex.}

\section{Conclusion}
Fast computational methods of the set of all PID controllers for linear continuous-time,
discrete-time and time-delay systems are proposed in this article. The driving force of the theory is the fact that non-convex stability regions can be built up easily by convex polygonal slices. The high computational speed results due to inspection of conditions at a relatively low number of (singular) frequencies. Only the results of the control group at DLR are surveyed here. A software tool called \robsin ~originated on that basis.

In author's opinion, the proposed design approach is especially elegant in the discrete-time domain, and of particular interest for time-delay systems. A powerful feature is the fact that, in principle, in all cases all design lines apply also when simultaneous stabilization of a set of plants is considered. This paves the basis for robust design of PID controllers. Indeed, while in all derivations of the article just a single representant is assumed, the extensions to the situation with a finite number of representants is straightforward. When applying the $k_P$-criterion, one would have to search for the intersection of stable $k_P-$intervals of each representant. And, inner-polygon candidates should provide the simultaneous stability for all representants. Yet some important issues, particularly those involving time-delay systems, remain open. For instance, it is not clear how to discriminate the frequency range with relevant singular frequencies for quasi-polynomials.


\end{document}